\documentclass[%
reprint,
superscriptaddress,
amsmath,amssymb,
aps,
]{revtex4-1}

\usepackage{color}
\usepackage{graphicx}
\usepackage{dcolumn}
\usepackage{bm}

\begin {document}

\title{Non-self-averaging behaviors and ergodicity in quenched trap models with finite system sizes
}

\author{Takuma Akimoto}
\email{takuma@rs.tus.ac.jp}
\affiliation{%
  Department of Physics, Tokyo University of Science, Noda, Chiba 278-8510, Japan
}%

\author{Eli Barkai}
\affiliation{%
  Department of Physics, Bar Ilan University, Ramat-Gan 52900, Israel
}%

\author{Keiji Saito}
\affiliation{%
  Department of Physics, Keio University, Yokohama, 223-8522, Japan
}%


\date{\today}

\begin{abstract}
Tracking tracer particles in heterogeneous environments plays an important role in unraveling  material properties. 
These heterogeneous structures are often static and depend on the sample realizations. Sample-to-sample 
fluctuations of such disorder realizations sometimes become considerably large.  
When we investigate the sample-to-sample fluctuations, fundamental averaging procedures are a thermal average for a single disorder 
realization and the disorder average for different disorder realizations. 
Here, we report on  non-self-averaging phenomena
in quenched trap models with finite system sizes, where we consider the periodic and the reflecting boundary conditions. 
Sample-to-sample fluctuations of diffusivity greatly exceed  
trajectory-to-trajectory fluctuations of diffusivity in the corresponding annealed model. 
For a single disorder realization, the time-averaged mean square displacement and position-dependent observables converge to constants
because of the existence of the equilibrium distribution. This is a manifestation  
of ergodicity.  As a result, the time-averaged quantities depend neither on the initial condition nor on the thermal histories but 
depend crucially on the disorder realization. 
\end{abstract}

\maketitle


\section{Introduction}
Elucidating anomalous dynamics in disordered media is of considerable theoretical and experimental interest 
\cite{Scher1975,bouchaud90,Mason1995,Yamamoto-Onuki-1998,metzler00,Havlin2002,Dauchot2005,Golding2006,Klafter2011}. 
In strongly disordered media such as a living cell, anomalous diffusion, i.e., a nonlinear growth of 
the mean square displacement (MSD), non-Gaussian propagator, and large sample-to-sample as well as trajectory-to-trajectory 
fluctuations of the MSDs are often observed \cite{Golding2006,Weigel2011,Jeon2011,Hofling2013,Tabei2013,Manzo2015}. 
In general, the fluctuations of long time-averaged observables are either a signature of ergodicity breaking or sample-to-sample  
variability, namely,  the effect of non-self averaging. 
The quenched trap model (QTM), which is a random walk model in a random potential landscape, 
is used to obtain a deep understanding of such anomalous dynamics \cite{bouchaud90, Bouchaud1992, Burov2011, Miyaguchi2011, Miyaguchi2015}, where 
the word ``quenched" implies that the random potential landscape does not change in time (more precisely, the time scale of changes of the 
random energy landscape is much larger than the time scale of the dynamics). 
In the QTM, there is a so-called glass temperature, below which many anomalous behaviors can be observed due to the divergence 
of the mean trapping time. In particular,  
the MSD grows as $\langle \{\bm{r}(t)- \bm{r}(0)\}^2 \rangle \propto t^\beta$ ($\beta < 1$), where $\bm{r}(t)$ is a position at time $t$. 
The power-law exponent $\beta$ characterizes anomalous diffusion and depends on the temperature as well as on
the space dimension \cite{bouchaud90}. 

When we treat statistical quantities in quenched environments,
there are basically three different averaging procedures to calculate the ensemble-averaged MSD,
i.e., thermal histories, initial conditions, and disorder realizations. Moreover, 
  another averaging procedure can be used to calculate the MSD in single-particle-tracking experiments \cite{kusumi1993,Golding2006}; e.g., 
one can use the time-averaged MSD defined by
\begin{equation}
 \label{tamsd_definition}
  \overline{\delta^{2}(\Delta;t)} \equiv
  \frac{1}{t - \Delta} \int_{0}^{t - \Delta} dt' \,
  [{\bm r}(t' + \Delta) - {\bm r}(t')]^{2} ,
\end{equation}
where $\Delta$ is the lag time, and $t$ is the measurement time.
For Brownian motions in homogeneous media, this time-averaged MSD as a function of the lag time $\Delta$ coincides with 
the ensemble-averaged MSD $\langle [{\bm r}(\Delta) - {\bm r}(0)]^2 \rangle$ if the measurement time is large and $\Delta \ll t$, where the 
average $\langle \cdot \rangle$ implies the thermal histories and the initial condition, which is uniform because of the homogeneity
of the environment.  
In other words, starting points do not play any role because the system is homogeneous. 

In highly heterogeneous systems, this equivalence will be broken  even when the local/temporal dynamics can be described by 
Brownian motion \cite{Massignan2014,AkimotoYamamoto2016,AkimotoYamamoto2016a}. 
In these systems, the time-averaged MSDs do not  coincide with the corresponding ensemble-averaged MSD. Moreover,  
the time-averaged MSD with a fixed $\Delta$ does not converge to a constant, but the trajectory-to-trajectory fluctuations are intrinsically random. 
Such intrinsic fluctuations of the time-averaged MSDs were found in the QTM \cite{Miyaguchi2011,Miyaguchi2015} as well as in
other stochastic models \cite{He2008,Metzler2014,Akimoto2013a}. These distributional behaviors imply either a breakdown of ergodicity 
or non-self averaging. In systems with a breakdown of ergodicity known as weak ergodicity breaking, distribution functions
 of time-averaged observables depend on the stochastic model as well as on the class of the observable \cite{Darling1957,Lamperti1958,Dynkin1961,He2008,Rebenshtok2007,Miyaguchi2011,Akimoto2013a}. 
 These distributional behaviors are related to 
a generalized concept of ergodicity, i.e., infinite ergodic theory \cite{Aaronson1997,Thaler1998,Akimoto2010,Akimoto2012,Akimoto2015}.

When a quenched disorder of a finite but large system is not so strong,  
the time-averaged MSDs will remain unchanged in different disorder realizations. In other words, the time-averaged MSDs 
do not strongly depend on the disorder realization.
This property is called self-averaging (SA) \cite{bouchaud90,Aharony1996}. 
When an observable in the system has an SA property, the observable does not depend on the disorder realization.
In the QTM, the quenched disorder is represented by the random energy landscape.
Because the diffusion coefficient in the system is determined by the mean jump rates (the inverse of the sample mean trapping time 
at the random potentials) \cite{Haus1987,Akimoto2016}, 
the SA property is a consequence of the law of large numbers for the mean trapping times at random potentials. 
Physically, this situation is related to the fact that a particle explores a large portion of the system, sampling many local environments, which 
is an approximate measure of the typical  disorder in the system \cite{Dentz2016,Russian2017}.
Therefore, the time-averaged MSD converges to a specific value, which is independent 
of the disorder realization when the system size (the number of random potentials) is sufficiently large. 
However, in the QTM below the glass temperature, the law of large numbers is broken 
due to the divergence of the mean trapping time. It is an interesting problem to clarify how the measured diffusivity and other 
observables under different disorder realizations spread in the large limit of the system size.

Continuous-time random walk (CTRW) is often used  
to understand anomalous dynamics in disordered media \cite{bouchaud90, metzler00, Klafter2011}.
It is an annealed model of the QTM and also a good approximation for the QTM 
when the dimension is greater than two \cite{Machta1985, Miyaguchi2011}. 
Although the CTRW has been extensively studied analytically and 
has successfully explained many aspects of anomalous diffusion in disordered media \cite{Scher1975, Shlesinger1982, Wong2004, He2008,Schulz2013}, 
it lacks the concept of disorder realization. Therefore, one cannot use the CTRW approximation when considering sample-to-sample 
fluctuations of disorder realizations. However, there are few exact  results for the QTM, unlike for CTRWs, 
because the quenched dynamics are crucially affected by 
the disorder realization. Therefore, it is important to obtain exact theoretical results for the randomness of the time-averaged MSD in the QTM. 
Moreover, it is interesting to compare fluctuations of time-averaged observables in the QTM with those in the CTRW, because the differences
between those fluctuations reveal unique dynamics of the QTM, which provide rich physical behaviors.

In this paper, we show how the quenched dynamics are different from the annealed ones by using the QTM with a finite system size and 
charactering the SA property by the SA parameter proposed in our previous study \cite{Akimoto2016}. 
Key facts that we use are ergodicity (existence of the equilibrium distribution for the particle's position)
and the generalized central limit theorem (stable law) for trapping times \cite{Feller1971}. Our main idea is to consider finite but large systems 
with disordered environments. With the aid of the finite system size, we rigorously obtain the equilibrium distribution of the particle's position, which
determines dynamical and static properties such as diffusivity and average particle position.
Using the key facts and considering the different disorder realizations, we provide a universal distribution 
of diffusivity below the glass temperature, which is a broad distribution, and thus the diffusivity is non-SA. 
We show that the sample-to-sample fluctuations of 
the time-averaged MSDs in different disorder realizations are substantially large compared with those in the annealed model (CTRW). 
A brief summary of a part of our results was recently published in Ref.~\cite{Akimoto2016}.

\section{model and equilibrium state}
As described above, we consider the QTM with a finite system size as a model of diffusion in strongly disordered environments.  
The QTM is a coarse-grained model of the diffusion in a continuous random energy landscape. 
The QTM that we consider here is a random walk in a quenched random energy landscape 
on a finite $d$-dimensional hypercubic lattice. 
Quenched disorder implies that  the disorder realization does not change with time during our measurement time scale.   
The lattice constant is set to unity, and the number of  lattices with different energies is finite.  Thus, 
a site ${\bm r}$ can be specified by ${\bm r}=(r_1, \cdots, r_d)$, e.g., ${ r}_k=l$  with $l=1,2, \cdots, L$ ($k=1, \cdots, d$). 
In the QTM, the random energy landscape is represented by the depths of the potential only. In other words, all the upper parts
of the potential energy landscape are exactly the same (see inset figures in Fig.~\ref{sts_msd}). We will soon discuss the boundary conditions. 

We assume that the depths of the potential energies at the sites are independent and identically distributed random variables. 
Moreover, we assume that the depth distribution $\rho (E)$ follows  an exponential distribution ($E>0$):
\begin{equation}
\rho(E) = T^{-1}_g \exp(-E/T_g), 
\label{exp_dist}
\end{equation}
where $T_g$ is a parameter called the glass temperature (here $k_B=1$). As will be shown, below the glass temperature, the mean of the trapping times in infinite systems
 diverges, and thus various anomalous behaviors in the dynamics can be observed \cite{bouchaud90, Monthus1996, Bertin2003, Monthus2003, Burov2007, Burov2011, Miyaguchi2011}. 
 A particle is trapped in the random potential and eventually escapes from the trap. It then jumps to one of the nearest neighbors with equal probability, 
 i.e., $1/(2d)$. 
The trapping time, that is, the time that  a particle is trapped in one of the valleys of the random potential landscape, is a random variable. 
According to the Arrhenius law and Kramers' results \cite{MELNIKOV1991}, 
the trapping-time distribution at site ${\bm r}$, $\rho (\tau; \tau_{\bm r})$,  follows the exponential distribution with  
 mean  $\tau_{\bm r} = \tau_0 \exp(E_{\bm r}/T)$, 
where $E_{\bm r}$ is the depth of the energy at site ${\bm r}$, 
$T$  the temperature, and $\tau_0$ a typical time scale. Because the depth of the energy $E_{\bm r}$ is generated by the exponential 
distribution, Eq.~(\ref{exp_dist}), $\tau_{\bm r}$ is also a random variable. 

Because we have  $L^d$ points on the lattice system, we use the set $\{ \tau_1 ,\cdots,   \tau_{L^d}\}$ for the mean trapping times
 (we used also $\tau_{\bm r}$, where the index ${\bm r}$ is a vector). 
For the given mean trapping times, i.e., $\tau_1, \cdots, \tau_{L^d}$, the probability density function (PDF) $\psi_L (\tau)$ of 
the trapping times in finite systems can be described by 
\begin{equation}
\psi_L (\tau) = \frac{1}{L^d} \sum_{j=1}^{L^d} \frac{1}{\tau_j} \exp(-\tau/\tau_j)
\label{pdf_L}
\end{equation}
in a mean-field sense (in a mean field, one uses a unique trapping-time PDF, so there is no spatial disorder). We called this PDF
 the sample mean PDF of the trapping times. If there is a correlation in the random energies, the sample mean PDF 
 becomes different form Eq.~(\ref{pdf_L}) \cite{Luo2018}. 
Using Eq.~(\ref{exp_dist}) and the continuous approximation, one can obtain 
 the PDF $\psi (\tau)$ of the trapping times in infinite systems as follows: 
 \begin{eqnarray}
 \psi (\tau) &=& \int_0^\infty \rho (\tau; \tau_{\bm r}) \rho (E)dE \nonumber\\
 &=& \int_0^\infty \exp \left( - \frac{\tau}{\tau_0} e^{-\frac{E}{T}} \right) \frac{1}{\tau_0 T_g}
 e^{ - \frac{E}{T_g}-\frac{E}{T} }dE.
 \end{eqnarray}
For large $\tau$ ($\gg 1$), $\exp \left( - \frac{\tau}{\tau_0} e^{-\frac{E}{T}} \right) $ can be ignored when $E$ is not sufficiently large; 
i.e., when $E \ll -T \ln (\tau_0/\tau)$. Therefore, one can approximate the integral as
\begin{equation}
 \psi (\tau) \propto \int_{-T \ln (\tau_0/\tau)}^\infty \frac{1}{\tau_0 T_g}
 e^{ - \frac{E}{T_g}-\frac{E}{T} } dE\propto \tau^{-1 - \alpha}
 \end{equation}
 for $\tau\to \infty$, where $\alpha \equiv T/T_g$. In what follows, we denote the PDF as $\psi_\alpha (\tau)$ 
 to express the explicit dependence on $\alpha$ and set the PDF as 
\begin{equation}
\int_\tau^\infty d\tau' \psi_\alpha (\tau') \sim \frac{\alpha c}{\Gamma(1-\alpha)} {\tau} ^{-\alpha}\quad (\tau \to \infty), 
\label{power-law-pdf}
\end{equation}
where $c$ is a constant that depends on $\tau_0$ and $\alpha$. 
We note that the mean trapping time, $\langle \tau \rangle = \int_0^\infty \tau \psi_\alpha (\tau)d\tau$, 
diverges for $T \leq T_g$. In infinite systems with quenched disorders, anomalous behaviors such as subdiffusion, 
weak ergodicity breaking, aging, and large sample-to-sample fluctuations are caused by the lack of a characteristic time scale (divergent 
mean trapping time)  \cite{Machta1985, bouchaud90, Monthus1996, Barkai2003aging, Bertin2003, Monthus2003, Burov2007, Burov2011, Miyaguchi2011, Metzler2014, Massignan2014, Miyaguchi2015, Luo2015}.
Due to the lack of the first moment, the Laplace transform of the PDF for $\alpha <1$ becomes the following form \cite{Bardou2002}:
\begin{equation}
\hat{\psi_\alpha}(s) \equiv \int_0^\infty \psi_\alpha (x) e^{-sx} dx = 1- c s^\alpha + o(s^\alpha),
\end{equation}
for $s\to 0$.

Because we consider finite systems with quenched disorders, the sample mean trapping time 
\begin{equation}
\mu_i (L) \equiv \frac{1}{L^d} \sum_{j=1}^{L^d} \tau_{j} 
\end{equation}
never diverges for all temperatures even when the temperature is below $T_g$, because  the number of sites in the sum is finite. 
With the aid of the finite characteristic time, the system can approach  equilibrium, and the equilibrium distribution is uniquely determined. 

The CTRW is an annealed model that mimics certain aspects of the dynamics of the QTM. In the CTRW, the particle jumps between nearest neighbors 
with waiting times drawn from Eq.~(\ref{power-law-pdf}), and the waiting-time distributions for all lattice points are identical. In that sense, the 
system is homogeneous and the CTRW is considered as the mean-field model of the QTM.
For infinite systems, statistical properties such as the squared displacement and  the number of jumps in the QTM with $d\geq 2$ 
can be approximately obtained by those in
the CTRW \cite{Machta1985}. However, for $d<2$, 
there are clear differences in the power-law exponent of the ensemble-averaged MSD and the scatter of the time-averaged MSDs 
 between the CTRW and the QTM \cite{bouchaud90, Miyaguchi2011, Miyaguchi2015}.  Therefore, 
 statistical laws for diffusivity depend on the dimension and whether the model is quenched or annealed for infinite systems.
 In the CTRW, the mean trapping time 
diverges even when the number of lattice points is finite. Thus, the system never reaches equilibrium, and 
shows weak ergodicity breaking \cite{Rebenshtok2007, Rebenshtok2008}.

Here, we consider two boundary conditions:  periodic and reflecting boundary conditions. 
The master equation for a single disorder realization $\tau_{\bm r}^{(i)}$ and $1<r_k<L$ ($k=1, \cdots, d$)
is given by
\begin{equation}
\frac{dP_{\bm r}}{dt} = \frac{1}{2d} \sum_{\bm r'}\frac{P_{\bm r'}}{\tau_{\bm r'}^{(i)}}  - \frac{P_{\bm r}}{\tau_{\bm r}^{(i)}},
\end{equation}
where the index $i$ represents a disorder realization, and 
the sum is  over the nearest neighbor sites and $P_{\bm r}$ is the probability of finding a particle at site ${\bm r}$. 
Here, $2 d$ is the number of nearest neighbors on the cubic lattices under consideration.
For the periodic boundary condition, the energies $E_{\bm r}$ are periodically arranged: $E_{r_k}=E_{r_{k+nL}}$ 
for all integers $k$ and $n$. For $d=1$, the master equation is given by
\begin{eqnarray*}
\frac{dP_{1}}{dt} = \frac{1}{2} \left(  
\frac{P_{2}}{\tau_{2}^{(i)}} + \frac{P_{L}}{\tau_{L}^{(i)}} \right) - \frac{P_{1}}{\tau_{1}^{(i)}}, \\
\frac{dP_{L}}{dt} = \frac{1}{2} \left(  
\frac{P_{L-1}}{\tau_{L-1}^{(i)}} + \frac{P_{1}}{\tau_{1}^{(i)}} \right) - \frac{P_{L}}{\tau_{L}^{(i)}}.
\end{eqnarray*}
For the reflecting boundary condition, a particle will return to the previous position when it hits  the boundary. 
For $d=1$
\begin{equation*}
\frac{dP_{1}}{dt} = \frac{1}{2} \left(  
\frac{P_{2}}{\tau_{2}^{(i)}} - \frac{P_{1}}{\tau_{1}^{(i)}} \right), \quad\frac{dP_{L}}{dt} = \frac{1}{2} \left(  
\frac{P_{L-1}}{\tau_{L-1}^{(i)}} - \frac{P_{L}}{\tau_{L}^{(i)}} \right).
\end{equation*}
A stationary solution (equilibrium state) in both cases is uniquely determined by
\begin{equation}
P_{\bm r}^{\rm eq} = \frac{\tau_{\bm r}^{(i)}}{L^d \mu_i (L)}.
\label{eq_state}
\end{equation}
We note that  Eq.~(\ref{eq_state}) is the exact solution for $\frac{dP_{\bm r}}{dt}=0$ for both boundary conditions. 
Because the disorder realization, i.e., $\tau_1, \cdots, \tau_{L^d}$, is completely different in different realizations, 
the equilibrium ensemble average crucially depends on the disorder realization. 
This equilibrium distribution can also be obtained by coarse graining of the continuous random energy landscape. 

\section{Non-self-averaging diffusivity and ergodicity for  the periodic boundary condition}
In this section, we focus on the periodic boundary condition, and consider sample-to-sample fluctuations (non-SA property) of 
the ensemble-averaged MSD and the ergodicity of the time-averaged MSD.

\subsection{Sample-to-sample fluctuations of the ensemble-averaged MSD}

Here, we consider the ensemble-averaged MSD, where the ensemble average is taken by many realizations of thermal histories and 
the initial condition (equilibrium distribution) for a fixed disorder realization. 
The MSD is determined by the mean number of jumps
and the second moment of the jump length in symmetric random walks. 
Therefore, the ensemble-averaged MSD for a fixed disorder realization is given by
  $\langle \{{\bm r}(t) - {\bm r}(0)\}^2 \rangle_{\rm eq} = \langle N_t \rangle_{\rm eq}$,
where $\langle N_t \rangle_{\rm eq}$ is the mean number of jumps until time $t$ and $\langle \cdot \rangle_{\rm eq}$ 
implies the equilibrium ensemble average, i.e., the initial points following the equilibrium state. 
Because the probability of finding a particle at ${\bm r}$ for the periodic boundary condition 
is invariant at equilibrium, the average  jump rate is also invariant. Using Eq.~(\ref{eq_state}) yields 
\begin{equation}
\sum_{\bm r} \frac{P_{\bm r}^{\rm eq}}{\tau_{\bm r}^{(i)}} = \frac{1}{\mu_i}, 
\end{equation}
where we deleted an $L$ dependence of $\mu_i(L)$.
Thus, $\langle N_t \rangle_{\rm eq}$ becomes 
\begin{equation}
\langle N_t \rangle_{\rm eq} = \frac{t}{\mu_i}
\label{renewal_function}
\end{equation}
for a  disorder realization $i$. This result is exact for any $t>0$ because the system is in equilibrium. 
Thus, the renewal function, i.e., the mean number of jumps as a function of time, 
is time-translation-invariant, i.e., $\langle N_t \rangle_{\rm eq} = \langle N_{t + t'} - N_{t'} \rangle_{\rm eq}$
for any $t'$ with the aid 
of the equilibration.  
We note that this average is taken over equilibrium initial conditions and thermal histories but 
not over disorder.  

Because there is no boundary in the sense that the position ${\bm r}(t)$ is not bounded, 
the MSD grows linearly with time; i.e., $\langle \{{\bm r}(t) - {\bm r}(0)\}^2 \rangle_{\rm eq} =t/\mu_i$ for any $t>0$. 
Hence, the diffusion coefficient defined by  $D_i \equiv \langle \{{\bm r}(t) - {\bm r}(0)\}^2 \rangle_{\rm eq}/t$ 
becomes 
\begin{equation}
D_i 
= \frac{1}{\mu_i}. 
\label{msd_mu}
\end{equation}
In equilibrium, the ensemble-averaged MSD is also time-translation-invariant; i.e., 
$\langle \{{\bm r}(t) - {\bm r}(0)\}^2 \rangle_{\rm eq} = \langle \{{\bm r}(t + t') - {\bm r}(t')\}^2 \rangle_{\rm eq}$ 
for any $t$ and $t'>0$.

 Here, we consider the SA property of the 
 ensemble-averaged MSD by taking the limit of $L\to \infty$.
 If the mean trapping time, which is given by $\langle \tau \rangle = \int_0^\infty \tau \psi_\alpha (\tau)d\tau$, 
 exists (does not diverge),  the law of large numbers holds for the sample mean of the trapping times. 
 Because the mean trapping time is finite  for $\alpha > 1$, 
we have
\begin{equation}
\frac{\tau_1^{(i)} + \cdots + \tau_{L^d}^{(i)}}{L^d} \rightarrow  \langle \tau \rangle \quad (L\to \infty),
\end{equation}
where $\tau_k^{(i)}$ is the mean trapping time at the $k$-th site for the $i$-th disorder realization. 
Because $\langle \tau \rangle$ is determined uniquely by the trapping-time distribution, 
the diffusion coefficient does not depend on the disorder realization.   
This is the SA property for diffusivity when $\alpha>1$.

\begin{figure}
\includegraphics[width=.9\linewidth, angle=0]{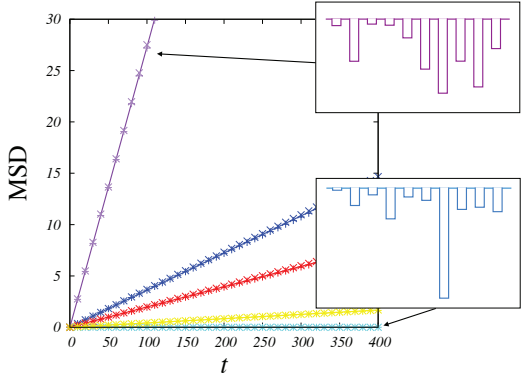}
\caption{Sample-to-sample fluctuations of the MSDs ($\alpha=0.5$ and $L=10$), where 
we consider five different disorder realizations. 
Cross  ($\times$) and plus ($+$) symbols are the results of the ensemble-averaged and the time-averaged MSDs, respectively.
Lines represent the theory, i.e., Eq.~(\ref{msd_mu}).  The initial points follow the equilibrium distributions in the ensemble-averaged MSDs 
and the measurement time is $t=10^7$ in the time-averaged MSD. 
The two disorder realizations (shapes of the random potentials) with the highest and the lowest diffusivities are shown in the inset figures.
 }
\label{sts_msd}
\end{figure}

On the other hand, the law of large numbers breaks down for $\alpha \leq 1$. Instead, the generalized 
central limit theorem holds for the  sum of $\tau_{k}^{(i)}$, which states that the
 PDF of the normalized sum of $\tau_{k}^{(i)}$, i.e., $\sum_{j=1}^{L^d} \tau_{j}^{(i)}/(L^d)^{1/\alpha}$,
follows the one-sided L\'evy distribution \cite{Feller1971}:
\begin{equation}
\frac{\tau_1^{(i)} + \cdots + \tau_{L^d}^{(i)}}{(L^d)^{1/\alpha}} \Rightarrow X_\alpha\quad (L\to \infty),
\label{gclt_1}
\end{equation}
where $X_\alpha$ is a random variable following the one-sided L\'evy distribution of index $\alpha$. 
More precisely, the Laplace transform of the PDF of $X_\alpha$, i.e., $\langle e^{-sX_\alpha}\rangle$, 
is given by 
\begin{equation}
\langle e^{-sX_\alpha}\rangle \sim e^{-cs^\alpha}.
\label{Laplace_Levy}
\end{equation}
Because the inverse Laplace transform of $e^{-s^\alpha}$ denoted by $\mathcal{L}^{-1} \{e^{-s^\alpha}\}(x)$ with $x>0$
 can be represented by the following infinite series \cite{Feller1971}:
\begin{equation}
\mathcal{L}^{-1} \{e^{-s^\alpha}\}(x) = -\frac{1}{\pi x} \sum_{k=1}^\infty \frac{\Gamma(k\alpha +1)}{k!} (-x^{-\alpha})^k \sin (k\pi \alpha),
\end{equation}
we have the PDF of $X_\alpha$ denoted by $l_\alpha(x)$ with $x>0$
\begin{equation}
l_\alpha(x) = -\frac{1}{\pi x} \sum_{k=1}^\infty \frac{\Gamma(k\alpha +1)}{k!} (-cx^{-\alpha})^k \sin (k\pi \alpha).
\label{levy_pdf}
\end{equation}
Using Eq.~(\ref{gclt_1}), the diffusion coefficient can be represented by 
\begin{equation}
D_i = L^{d(1-1/\alpha)} X_\alpha^{-1}.
\end{equation}
Because the PDF of $X_\alpha$ is not a delta function, $D_i$ has sample-to-sample fluctuations; i.e., 
it depends crucially  on the disorder realization. In Fig.~\ref{sts_msd}, we plot both the ensemble-averaged and the time-averaged 
MSDs.  The two MSDs almost coincide because of the ergodicity, which will be shown later. 

\begin{figure}
\includegraphics[width=.6\linewidth, angle=-90]{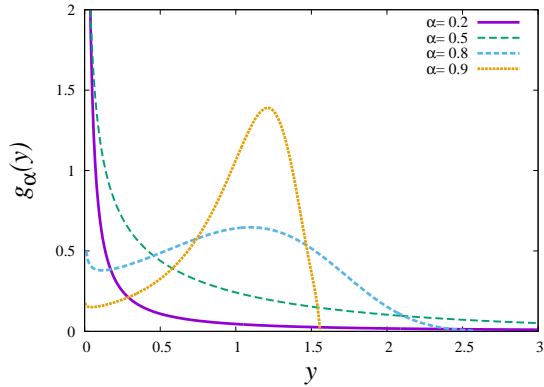}
\caption{Inverse L\'evy PDF used to describe the fluctuations of MSD in the QTM, for different $0<\alpha=T/T_g<1$
All PDFs are unbounded at the origin. 
 }
\label{inv_Levy_dist}
\end{figure}

\begin{figure}
\includegraphics[width=.7\linewidth, angle=-90]{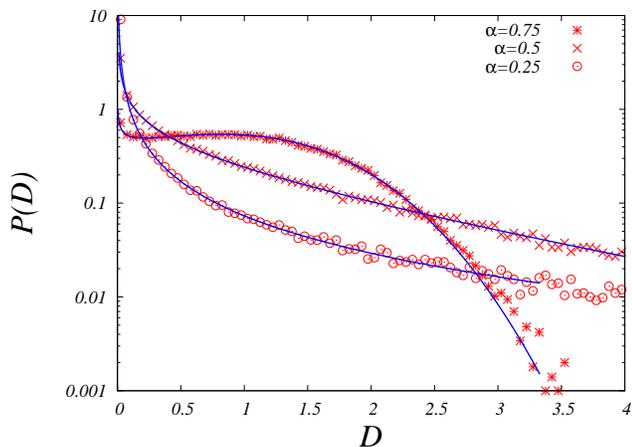}
\caption{Probability density functions of $D_i$ for different $\alpha$, where we consider $10^5$ different disorder realizations.
Symbols are the results of numerical 
simulations ($d=1$ and $L=10^5$), and the solid curves are the theoretical results.
 }
\label{dist_D_alpha}
\end{figure}

The PDF of $X_\alpha^{-1}$ can be explicitly represented 
by using the one-sided  L\'evy distribution:  
\begin{equation}
\Pr (X_\alpha^{-1} \leq y) = \Pr (X_\alpha \geq y^{-1}) = \int_{y^{-1}}^\infty l_\alpha (x)dx.
\label{inverse_Levy_cum}
\end{equation}
We call this distribution the inverse L\'evy distribution.
Differentiating Eq.~(\ref{inverse_Levy_cum}), we obtain the PDF of $X_\alpha^{-1}$, denoted by $g_\alpha(y)$:
\begin{equation}
g_\alpha(y) = -\frac{1}{\pi y} \sum_{k=1}^\infty \frac{\Gamma(k\alpha +1)}{k!} (-cy^{\alpha})^k \sin (k\pi \alpha).
\label{PDF_inverse_Levy}
\end{equation}
The inverse L\'evy distribution is a special one of
  the modified Mittag-Leffler distribution, which is a one-parameter extension of the Mittag-Leffler distribution \cite{Miyaguchi2015}.
The PDFs of the inverse L\'evy distributions for different exponents $\alpha$ are represented in Fig.~\ref{inv_Levy_dist}. 
From Eq.~(\ref{PDF_inverse_Levy}), we have $g_\alpha(y) \propto y^{\alpha-1}$ for $y\to 0$. In other words, the PDF
 is unbounded at the origin, corresponding to very small diffusivity in some disorder realizations, which cannot be observed 
in the annealed model (CTRW). Figure~\ref{dist_D_alpha} shows the PDF of $D_i$ obtained by numerical simulations, where
we generated the random energy and calculated the sample average $\mu_i$ to obtain $D_i$ by $D_i=1/\mu_i$.
The inverse L\'evy distribution is the exact distribution of $D_i$ for any dimension and  is
valid for finite and large $L$. 
 
Here, we derive the Laplace transform of the inverse L\'evy PDF (\ref{PDF_inverse_Levy}). In the same way as in Ref.~\cite{Miyaguchi2011}, 
we use the auxiliary distribution: $G_\alpha(y,h) \equiv \Pr (X_\alpha > h y^{-1})$. The Laplace transform of 
$G_\alpha(y,h)$ with respect to $h$ is given by
\begin{equation}
\hat{G}_\alpha (y,s) \equiv \int_0^\infty dh e^{-sh} G_\alpha(y,h) = \frac{1-e^{-cs^\alpha y^\alpha}}{s},
\end{equation}
where we used the Laplace transform of the one-sided L\'evy distribution, i.e., Eq.~(\ref{Laplace_Levy}). 
Moreover, the Laplace transform with respect to $y$ gives 
\begin{align}
\hat{G}_\alpha (\nu,s) &\equiv \int_0^\infty e^{-\nu x} G_\alpha (y,s) dy\nonumber\\
&= \frac{1}{s} \left[ \frac{1}{\nu} -  \sum_{k=0}^\infty \frac{(-\nu)^k \Gamma(\frac{k+1}{\alpha}) (cs^\alpha)^{-\frac{k+1}{\alpha}}}{\alpha k!}
\right].
\end{align}
The inverse Laplace transform with respect to $s$ gives
\begin{equation}
\hat{G}_\alpha (\nu,h) =\frac{1}{\nu} -  \sum_{k=0}^\infty \frac{(-\nu)^k \Gamma(\frac{k+1}{\alpha}) }{\alpha k! \Gamma(k+2)}
\left(\frac{h^\alpha}{c} \right)^{\frac{k+1}{\alpha}}.
\end{equation}
Hence, the Laplace transform of the inverse L\'evy PDF is given by 
\begin{equation}
\hat{g}_\alpha (\nu) = \nu \hat{G}_\alpha (\nu,1) 
=1 -  \sum_{k=0}^\infty \frac{(-1)^k \nu^{k+1} \Gamma(\frac{k+1}{\alpha}) }{\alpha k! (k+1)!}
\left(\frac{1}{c} \right)^{\frac{k+1}{\alpha}}.
\end{equation}
Using the relation between the Laplace transform and the moments, we have the first and second moments of $X^{-1}_\alpha$:
\begin{equation}
\langle X_\alpha^{-1} \rangle = \frac{\Gamma(\frac{1}{\alpha})}{\alpha c^{\frac{1}{\alpha}}},
\quad \langle X_\alpha^{-2} \rangle = \frac{\Gamma(\frac{2}{\alpha})}{\alpha c^{\frac{2}{\alpha}}}.
\label{moments_inv_Levy}
\end{equation}
From Eq.~(\ref{moments_inv_Levy}),  
we obtain the disorder average of $D_i$:
\begin{equation}
\langle D \rangle_{\rm dis} = \frac{L^{d(1-1/\alpha)}\Gamma(\alpha^{-1})}{\alpha c^{1/\alpha}},
\label{D_dis_ave}
\end{equation}
where $\langle \cdot \rangle_{\rm dis}$ means the disorder average. 
Recall that 
$\langle D \rangle_{\rm dis}$ has  units of m${}^2$/s.

\begin{figure}
\includegraphics[width=.6\linewidth, angle=-90]{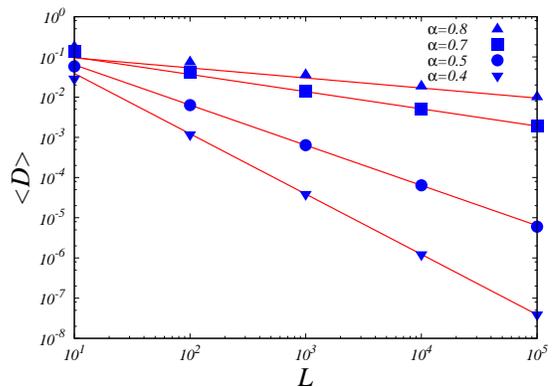}
\caption{Disorder average of the diffusion coefficients as a function of the system size $L$ for several $\alpha$ ($d=1$), 
where we used $10^5$ different disorder realizations, and the disorder average $\langle D \rangle_{\rm dis}$ can be calculated 
by taking the disorder average of $D_i=1/\mu_i$. 
Here, the symbols are the results of numerical simulations, and   
the solid lines represent Eq.~(\ref{D_dis_ave}).
 }
\label{disave_D}
\end{figure}

The disorder average of $D_i$ depends on $L$ and thus  becomes zero as  $L$ goes to infinity:
$\langle D \rangle_{\rm dis} =  L^{d(1-1/\alpha)} \langle X_\alpha^{-1} \rangle_{\rm dis}\to 0$ as $L\to \infty$ (see Fig.~\ref{disave_D}). 
In infinite systems, the MSD grows as $\langle \{{\bm r}(t) - {\bm r}(0)\}^2 \rangle
\propto t^{\beta}$ with $\beta<1$. Because the diffusion coefficient can be defined as the limit for the slope of the MSD, i.e., 
${\displaystyle D\equiv \lim_{t\to\infty} \langle \{{\bm r}(t) - {\bm r}(0)\}^2 \rangle}/t$, it becomes zero in infinite systems, which 
is consistent with the fact that $\langle D \rangle_{\rm dis} \to 0$ as  $L\to \infty$.

\begin{figure}
\includegraphics[width=.7\linewidth, angle=-90]{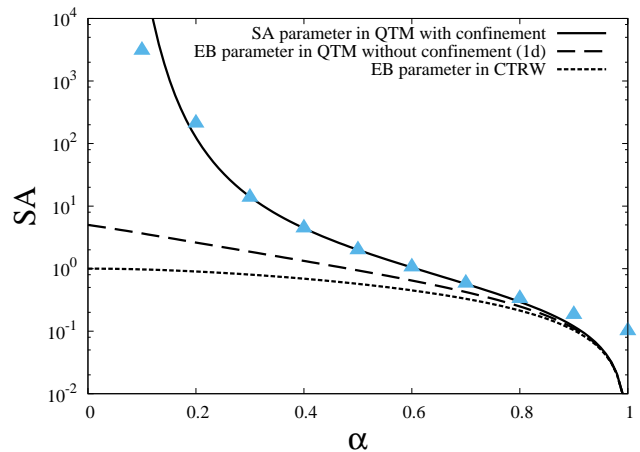}
\caption{Self-averaging (SA) parameter as a function of $\alpha$. The symbols are the result of a numerical 
simulation ($d=1$ and $L=10^3$), where we used $10^5$ different disorder realizations. 
The solid line represents the theory, Eq. (\ref{SA_theory}). The dashed and dotted lines represent the EB parameters 
in the QTM with no confinement and the CTRW, respectively \cite{Miyaguchi2011, He2008}. 
 }
\label{SA_L=1000}
\end{figure}

Because the diffusion coefficients exhibit sample-to-sample fluctuations, we quantify the non-SA property by the SA parameter, 
defined by
\begin{equation}
{\rm SA}(L; D)\equiv \frac{\langle D^2 \rangle_{\rm dis}  - \langle D \rangle_{\rm dis}^2}
{\langle D \rangle_{\rm dis}^2 }.
\label{SA}
\end{equation} 
If the SA parameter becomes zero for $L\to \infty$, the system is called SA because sample-to-sample fluctuations become zero 
 when the systems become large. 
Using the first and second moment of $D_i$, we have 
\begin{equation}
\lim_{L\to \infty}  {\rm SA}(L;D) 
=
\left\{
\begin{array}{ll}
0 &(\alpha >1)\\
\\
\dfrac{\alpha \Gamma(\frac{2}{\alpha}) }{\Gamma(\frac{1}{\alpha})^2} -1\quad &(\alpha \leq 1).
\end{array}
\right.
\label{SA_diffusivity}
\end{equation}
It follows that the diffusion coefficient is not SA for $\alpha<1$. The SA parameter becomes exponentially larger than the ergodicity
breaking (EB) parameter defined below 
for the corresponding infinite system especially for small $\alpha$ (see Fig.~\ref{SA_L=1000}).

\subsection{ Ergodicity of the time-averaged MSD}
Here, we consider the trajectory-to-trajectory fluctuations of the time-averaged MSD for a fixed disorder realization, 
i.e., the ergodic property of the time-averaged MSD. If the time-averaged MSD is an ergodic observable, it converges to 
a constant in the long-time limit. In other words, there are no intrinsic fluctuations in the time-averaged MSDs if the system is ergodic. 
To characterize the ergodic property quantitatively, we use the EB parameter \cite{He2008} defined by 
\begin{equation}
{\rm EB}(t;\Delta)\equiv \frac{\langle \{\overline{\delta^{2}(\Delta;t)}\}^2 \rangle  - \langle \overline{\delta^{2}(\Delta;t)} \rangle^2}
{\langle \overline{\delta^{2}(\Delta;t)} \rangle^2},
\label{EB}
\end{equation} 
where $\langle \cdot \rangle$ implies an average not only with respect to initial conditions, 
but also with respect to thermal histories. In what follows, we consider 
the equilibrium initial ensemble. We note that the disorder realization  is unique while the thermal histories and the initial conditions are different.
If the EB parameter goes to zero for $t\to \infty$, the time-averaged MSDs for a single disorder realization 
converge to a constant, which depends on neither the thermal histories nor on the initial point. The EB parameter is  
used not only to investigate the ergodic properties but also to extract the underlying information on the dynamics 
\cite{Akimoto2011,Uneyama2012,Uneyama2015}.

Because MSD with the equilibrium ensemble is time-translation-invariant, the ensemble average of the time-averaged MSD 
is given by
\begin{eqnarray}
\langle \overline{\delta^{2}(\Delta;t)} \rangle_{\rm eq} &=& \frac{1}{t - \Delta} \int_{0}^{t - \Delta} dt' \,
\langle  [{\bm r}(t' + \Delta) - {\bm r}(t')]^{2} \rangle_{\rm eq} \nonumber\\
&=& \frac{\Delta}{\mu_i}.
\end{eqnarray}
Here, we calculate the second moment of the time-averaged MSD. We assume $\Delta \ll t$. We then have
\begin{widetext}
\begin{align}
\langle \{\overline{\delta^{2}(\Delta;t)}\}^2 \rangle_{\rm eq} 
\sim \frac{2}{t^2} \int_0^t dt' \int_{t'}^t dt'' \langle [{\bm r}(t' + \Delta) - {\bm r}(t')]^{2} [{\bm r}(t'' + \Delta) - {\bm r}(t'')]^{2} \rangle_{\rm eq}.
\end{align}
Moreover, we assume that the displacements ${\bm r}(t' + \Delta) - {\bm r}(t')$ and ${\bm r}(t'' + \Delta) - {\bm r}(t'')$ ($t''>t'$) are 
independent if $t'' > t' +\Delta$. This assumption is not exact in general. However, it does not affect the following result crucially.
\begin{align}
\langle \{\overline{\delta^{2}(\Delta;t)}\}^2 \rangle_{\rm eq} 
&= \frac{2}{t^2} \int_0^{t-\Delta} dt' \int_{t'}^{t'+\Delta} dt'' \langle [{\bm r}(t' + \Delta) - {\bm r}(t')]^{2} [{\bm r}(t'' + \Delta) - {\bm r}(t'')]^{2} \rangle_{\rm eq}
\nonumber\\
&+ \frac{2}{t^2} \int^t_{t-\Delta} dt' \int_{t'}^{t} dt'' \langle [{\bm r}(t' + \Delta) - {\bm r}(t')]^{2} [{\bm r}(t'' + \Delta) - {\bm r}(t'')]^{2} \rangle_{\rm eq}
\nonumber\\
&+ \frac{2}{t^2} \int_0^{t-\Delta} dt' \int_{t'+\Delta}^t dt'' \langle [{\bm r}(t' + \Delta) - {\bm r}(t')]^{2} \rangle_{\rm eq} 
\langle [{\bm r}(t'' + \Delta) - {\bm r}(t'')]^{2} \rangle_{\rm eq}.
\end{align}
Dividing the displacements ${\bm r}(t' + \Delta) - {\bm r}(t')$ and ${\bm r}(t'' + \Delta) - {\bm r}(t'')$ ($t''>t'$) into 
${\bm r}(t' + \Delta)  - {\bm r}(t'')+ {\bm r}(t'') - {\bm r}(t')$ and ${\bm r}(t'' + \Delta) - {\bm r}(t' + \Delta)  + {\bm r}(t' + \Delta) - {\bm r}(t'')$, 
we have 
\begin{align}
\langle \{\overline{\delta^{2}(\Delta;t)}\}^2 \rangle_{\rm eq} 
&= \frac{2}{t^2} \int_0^{t-\Delta} dt' \int_{t'}^{t'+\Delta} dt''\left\{ \langle [{\bm r}(t'') - {\bm r}(t')]^2 \rangle_{\rm eq} 
\langle [{\bm r}(t'' + \Delta) - {\bm r}(t'')]^2 \rangle_{\rm eq} 
+ \langle [{\bm r}(t' + \Delta) - {\bm r}(t'')]^{4} \rangle_{\rm eq} \right. \nonumber\\
&+ \langle [{\bm r}(t' + \Delta) - {\bm r}(t'')]^2 \rangle_{\rm eq} \langle [{\bm r}(t'' + \Delta) - {\bm r}(t' + \Delta)]^2 \rangle_{\rm eq} \} \nonumber\\
&+ \frac{2}{t^2} \int^t_{t-\Delta} dt' \int_{t'}^{t} dt''\left\{ \langle [{\bm r}(t'') - {\bm r}(t')]^2 \rangle_{\rm eq} 
\langle [{\bm r}(t'' + \Delta) - {\bm r}(t'')]^2 \rangle_{\rm eq} 
+ \langle [{\bm r}(t' + \Delta) - {\bm r}(t'')]^{4} \rangle_{\rm eq} \right.\nonumber\\
&+ \langle [{\bm r}(t' + \Delta) - {\bm r}(t'')]^2 \rangle_{\rm eq} \langle [{\bm r}(t'' + \Delta) - {\bm r}(t' + \Delta)]^2 \rangle_{\rm eq} \}\nonumber\\
&+\frac{2}{t^2} \int_0^{t-\Delta} dt' (t-t'-\Delta)\langle N_\Delta \rangle_{\rm eq}^2\nonumber\\
&= \frac{2}{t^2} \int_0^{t-\Delta} dt' \int_{t'}^{t'+\Delta} dt''\left\{\langle N_{t'' - t'} \rangle_{\rm eq} \langle N_{\Delta} \rangle_{\rm eq} 
+ \left(1+\frac{2}{d} \right) \langle N_{t' + \Delta - t''}^2 \rangle_{\rm eq} - \frac{2}{d} \langle N_{t' + \Delta - t''} \rangle_{\rm eq} \right. \nonumber\\
&\left.  + \langle N_{t' + \Delta - t''} \rangle_{\rm eq} \langle N_{t'' - t'} \rangle_{\rm eq}  \right\} \nonumber\\
&+ \frac{2}{t^2} \int^t_{t-\Delta} dt' \int_{t'}^{t} dt''\left\{\langle N_{t'' - t'} \rangle_{\rm eq} \langle N_{\Delta} \rangle_{\rm eq} 
+ \left(1+\frac{2}{d} \right)  \langle N_{t' + \Delta - t''}^2 \rangle_{\rm eq} - \frac{2}{d} \langle N_{t' + \Delta - t''} \rangle_{\rm eq} \right. \nonumber\\
&\left.  + \langle N_{t' + \Delta - t''} \rangle_{\rm eq} \langle N_{t'' - t'} \rangle_{\rm eq}  \right\} 
+\left(1 - \frac{\Delta}{t}\right)^2 \left( \frac{\Delta}{\mu_i}\right)^2 \nonumber\\
&= \left(\frac{\Delta}{\mu_i}\right)^2 + \frac{2\Delta^2}{d t} \left(\frac{2\Delta}{3 \mu_i^2} -  \frac{A\mu_i - 1}{ \mu_i} \right) +o(t^{-1}),
\label{EB-confined}
\end{align}
\end{widetext}
where we used 
\begin{equation}
\langle [{\bm r}(t) - {\bm r}(0)]^4 \rangle_{\rm eq} =  \left(1+\frac{2}{d} \right) \langle N_t^2 \rangle_{\rm eq} - \frac{2}{d}
\langle N_t \rangle_{\rm eq},
\end{equation}
whose derivation depends on the fact that displacements are determined by the number of jumps and jumps are homogeneous 
(with equal probability); a mean-field approximation is used to obtain the second moment of $N_t$, i.e., 
$\langle N_t^2\rangle_{\rm eq} = (t/\mu_i)^2 + At$, where $A$ is a constant (see Appendix~A).

\begin{figure}[b]
\includegraphics[width=.7\linewidth, angle=-90]{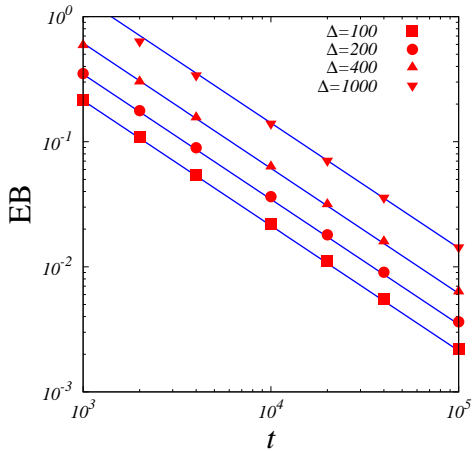}
\caption{Ergodicity breaking (EB) parameters as a function of the measurement time in a single disorder realization ($d=1, T=1, T_g=1.5$, and $L=10$),
where the initial position follows the equilibrium ensemble, and the boundary condition is periodic. 
Lines represent the theoretical results, Eq.~(\ref{EB_qtm}), for different $\Delta$, where the fitting parameter $B$ does not depend on
$\Delta$ ($B=60$).}
\label{EB_T=1_Tg=1.5}
\end{figure}

As a result, the EB parameter for a single disorder realization decays as
\begin{equation}
{\rm EB}(t;\Delta) \sim \frac{4\Delta}{3dt} \left(1+ \frac{B}{\Delta} \right) \quad (t \to \infty),
\label{EB_qtm}
\end{equation} 
where $B$ is a constant that depends on the disorder realization as well as $\mu_i$ but not on $\Delta$. In fact, 
the EB parameter decays as Eq.~(\ref{EB_qtm}) (see Fig.~\ref{EB_T=1_Tg=1.5}). 
This $1/t$ scaling is universal for ergodic systems even when the environment is not homogeneous \cite{Miyaguchi2017}. 
Because the variance of the time-averaged MSD goes to zero, the time-averaged MSD converges to a constant:
\begin{equation}
\overline{\delta^{2}(\Delta;t)} \to \langle \{{\bm r} (\Delta) - {\bm r}(0)\}^2 \rangle_{\rm eq}
\end{equation}
for $t\to \infty$. 
This statement becomes invalid for an infinite system ($L=\infty$) because 
there is no equilibrium state in that case. Although we assume the equilibrium initial condition, the EB parameter goes to 
zero even without an initial equilibrium condition. This is because all time-averaged MSDs starting from different initial positions 
converge to the same constant, i.e., $\langle \{{\bm r} (\Delta) - {\bm r}(0)\}^2 \rangle_{\rm eq}$. 

\begin{figure}
\includegraphics[width=.9\linewidth, angle=0]{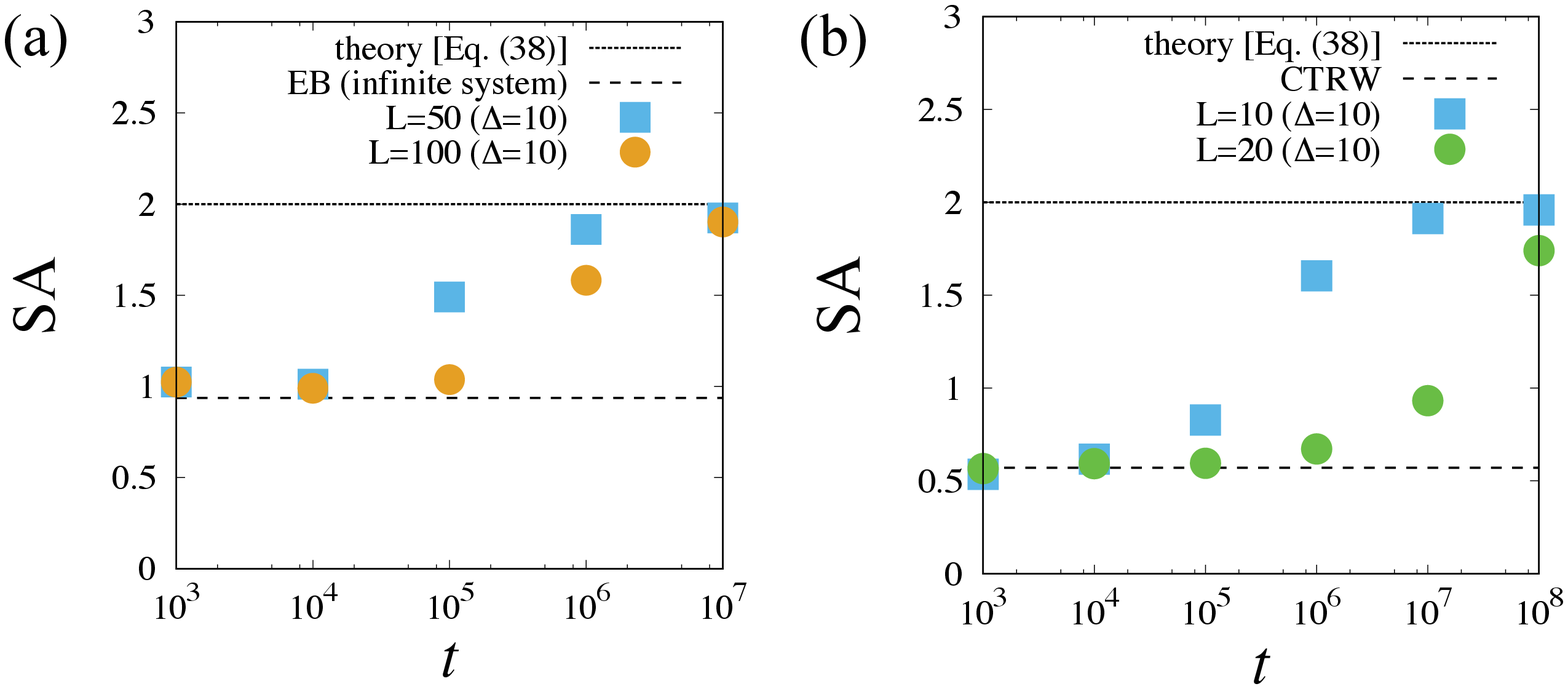}
\caption{Self-averaging (SA) parameter as a function of the measurement time ($\Delta =10$), where the initial points are chosen 
by the equilibrium distribution, 
and the boundary condition is periodic ($\alpha=0.5$). 
Symbols represent the results of numerical 
simulations for (a) $d=1$ with different $L$, and the dashed line is the EB parameter
in the QTM with no confinement \cite{Miyaguchi2011}; (b) $d=3$ with different  $L$, and the dashed line is the EB parameter in the QTM 
with no confinement  \cite{Miyaguchi2011}, which is the same as that of the CTRW 
 \cite{He2008}. The time when the SA parameter converges to the theoretical value crucially depends on $L$.   
 }
\label{SA_t-dependence}
\end{figure}

Thus far, sample-to-sample fluctuations of the diffusivity have been characterized by the asymptotic behavior of the SA parameter, i.e., 
 ${\displaystyle \lim_{L\to \infty} {\rm SA} (L;D)}$. 
Here, we define the SA parameter for the time-averaged MSD as a function of $t$ and $L$:
\begin{align}
 {\rm SA}(t,L;\delta{\bm r}^2_\Delta) 
= \frac{\langle \overline{\delta^{2}(\Delta;t)}^2 \rangle_{\rm dis}  - \langle \overline{\delta^{2}(\Delta;t)}\rangle_{\rm dis}^2}
{\langle \overline{\delta^{2}(\Delta;t)}\rangle_{\rm dis}^2 },
\end{align}
where $\delta {\bm r}_\Delta \equiv {\bm r}(t+\Delta) - {\bm r}(t)$. Because the time-averaged MSD is ergodic for finite $L$, 
taking the long-time limit gives 
\begin{align}
{\rm SA}(t,L;\delta{\bm r}^2_\Delta) 
\to \frac{\langle 1/\mu_i^2 \rangle_{\rm dis}  - \langle 1/\mu_i \rangle_{\rm dis}^2}{\langle 1/\mu_i \rangle_{\rm dis}^2 } 
\quad (t\to\infty).
\end{align}
Furthermore, taking the large-$L$ limit, we can characterize the sample-to-sample fluctuations in the time-averaged MSD:
\begin{equation}
\lim_{L\to \infty} \lim_{t\to \infty} {\rm SA}(t,L;\delta{\bm r}^2_\Delta) 
=
\left\{
\begin{array}{ll}
0 &(\alpha >1)\\
\\
\dfrac{\alpha \Gamma(\frac{2}{\alpha}) }{\Gamma(\frac{1}{\alpha})^2} -1\quad &(\alpha \leq 1).
\end{array}
\right.
\label{SA_theory}
\end{equation}
This is exactly the same as the SA parameter of the diffusivity, i.e., Eq.~(\ref{SA_diffusivity}).
Figure~\ref{SA_t-dependence} shows the $t$ dependence of the SA parameter. When the measurement time $t$ 
is not sufficiently large, particles do not explore the whole space and rarely hit the boundary. Therefore, sample-to-sample 
fluctuations of diffusivity in this regime are similar to those in the system with no confinement. After exploring almost the whole 
region, the SA parameter gradually approaches the theoretical value.

\section{Non-self-averaging properties for the reflecting boundary condition}
Here, we consider the reflecting boundary condition. 
We investigate the confinement effect for the MSD and sample-to-sample fluctuations of the average position.

\subsection{Sample-to-sample fluctuations of the MSD}
Due to the confinement, the MSD typically exhibits two different behaviors. For small $t$ and $L\gg 1$, 
it increases linearly with time. Approximately, it becomes 
 \begin{equation}
 \langle \{{\bm r}(t) - {\bm r}(0)\}^2 \rangle_{\rm eq} \sim \langle N_t \rangle_{\rm eq}
 \end{equation}
  because particles do not encounter the boundary
 in this regime. On the other hand, it becomes a constant for large $t$ because of the confinement. 
 Because the system is in equilibrium, both ${\bm r}(t)$ and ${\bm r}(0)$ follow the equilibrium distribution and become independent  
of $t$ in the long-time limit. The MSD becomes 
 \begin{equation}
 \langle \{{\bm r}(t) - {\bm r}(0)\}^2 \rangle_{\rm eq} \to 2(\langle {\bm r}^2 \rangle_{\rm eq} 
 - \langle {\bm r} \rangle_{\rm eq}^2)
 \end{equation}
  in the long-time limit. We will focus on the fluctuations of  $2\sigma_i^2 = 2(\langle {\bm r}^2 \rangle_{\rm eq} - \langle {\bm r} \rangle_{\rm eq}^2)$, 
  where the index $i$ represents a disorder realization.
Here, we define the crossover time $t_c$ from the diffusive to the plateau regime as the time when $\langle N_{t_c} \rangle = \sigma_i^2$,  
which is given by $t_c= 2\mu_i \sigma_i^2$. Because $\mu_i$ and $\sigma_i^2$ depend crucially on the disorder realization, 
$t_c$ also depends on the disorder realization. Note that the disorder average of $t_c$ always diverges for $L>1$ because of 
the divergence of $\langle \mu_i \rangle_{\rm dis}$.
 
 For the first regime, i.e., $t \ll t_c$, the MSD grows almost linearly with time and the diffusion coefficient in this regime 
 can be approximated as $1/\mu_i$. 
 Therefore, the disorder average and sample-to-sample fluctuations of the MSD in the first regime are almost the same as 
 those in the case of the periodic boundary. In particular, the disorder average of the diffusion coefficient as a function of time, i.e., 
 Eq.~(\ref{D_dis_ave}), and the asymptotic behavior of the SA parameter, i.e., Eq.~(\ref{SA}), are valid in this regime. 
 
 For the plateau regime, i.e., $t\gg t_c$, the MSD becomes a constant: $\langle \{{\bm r}(t) - {\bm r}(0)\}^2 \rangle_{\rm eq} \sim 
 2\sigma^2_i$. To consider the disorder average of $\sigma_i$, we derive the distribution of observables $\mathcal{O}$ that depend 
 only on the position. The following calculation is almost the same as in Refs.~\cite{Rebenshtok2007, Rebenshtok2008}.
 We denote it by $\mathcal{O}_{\bm r}$. The equilibrium ensemble average can be represented by
 \begin{equation}
 \langle \mathcal{O} \rangle_{\rm eq} = \sum_{\bm r}  \mathcal{O}_{\bm r} P_{\bm r}^{\rm eq}.
 \end{equation}
Let $f_\alpha (\mathcal{O})$ be the PDF of $\langle \mathcal{O} \rangle_{\rm eq}$; we then  
  have
  \begin{eqnarray}
  f_\alpha (\mathcal{O}) &=& \left\langle \delta \left(  \mathcal{O}  -  \sum_{\bm r}  \mathcal{O}_{\bm r} P_{\bm r}^{\rm eq}\right) \right\rangle_{\rm dis}
  \nonumber\\
  &=& -\frac{1}{\pi} \lim_{\varepsilon \to 0} {\rm Im} \left\langle   \frac{1}{ \mathcal{O}  +i\varepsilon -  \sum_{\bm r}  \mathcal{O}_{\bm r} P_{\bm r}^{\rm eq} }
  \right\rangle_{\rm dis}
  \nonumber\\
  &=& -\frac{1}{\pi} \lim_{\varepsilon \to 0} {\rm Im} \frac{1}{\mathcal{O} + i \varepsilon} 
  \left\langle   \frac{1}{ 1 - \frac{1}{\mathcal{O}  +i\varepsilon}   \sum_{\bm r}  \mathcal{O}_{\bm r} P_{\bm r}^{\rm eq} }
  \right\rangle_{\rm dis}.
  \end{eqnarray}
We note that $\langle \mathcal{O} \rangle_{\rm eq}$ is a random variable that depends on the disorder realization. 
Using the generating function, given by 
\begin{equation}
\hat{f}_\alpha (\xi) \equiv \sum_{k=0}^\infty (-1)^k \langle \langle \mathcal{O} \rangle_{\rm eq}^k \rangle_{\rm dis} \xi^k
= \left\langle \frac{1}{1+\xi  \langle \mathcal{O} \rangle_{\rm eq}} \right\rangle_{\rm dis},
\end{equation}
we have
\begin{equation}
f_\alpha (\mathcal{O}) = -\frac{1}{\pi} \lim_{\varepsilon \to 0}{\rm Im} \frac{1}{\mathcal{O} + i \varepsilon} \hat{f}_\alpha 
\left(-\frac{1}{\mathcal{O} + i \varepsilon }\right).
\end{equation}
Using Eq.~(\ref{eq_state}), we obtain
\begin{widetext}
\begin{align}
\hat{f}_\alpha (\xi) &= \left\langle \int_0^\infty ds e^{- (1+\xi  \sum_{\bm r} \mathcal{O}_{\bm r} \tau_{\bm r}/t_{L})s} \right\rangle_{\rm dis} \nonumber\\
&= \int_0^\infty ds \int_0^\infty dt_{L} \int_0^\infty d\tau_1 \psi_\alpha(\tau_1) \cdots \int_0^\infty d\tau_{L^d} \psi_\alpha (\tau_{L^d})
e^{- (1+\xi  \sum_{\bm r} \mathcal{O}_{\bm r} \tau_{\bm r}/t_{L})s} \delta (t_L - \sum_{\bm r} \tau_{\bm r}),
\end{align}
where $t_L \equiv \sum_{\bm r} \tau_{\bm r}$. Here, we approximate $\psi_\alpha (\tau)$ as a stable distribution with exponent $\alpha$:
\begin{equation}
\int_0^\infty \psi_\alpha (\tau_{\bm r}) e^{-s\tau_{\bm r}}dt\tau_{\bm r} = \exp (-cs^\alpha).
\label{Laplace-stable}
\end{equation}
Using Eq.~(\ref{Laplace-stable}) and the Fourier representation of the delta function gives 
\begin{align}
\hat{f}_\alpha (\xi) 
= \int_0^\infty ds \int_0^\infty dt_{L} \int_0^\infty \frac{dk}{2\pi t_L} 
\exp\left[- ik -s -  c \sum_{\bm r}  (ik + \mathcal{O}_{\bm r}\xi s)^\alpha/t_{L}^\alpha \right].
\end{align}
\end{widetext}
Using the same technique given in Ref.~\cite{Rebenshtok2008}, we have
\begin{equation}
\hat{f}_\alpha (\xi) = \frac{\sum_{\bm r} (1+\mathcal{O}_{\bm r} \xi)^{\alpha-1}}{\sum_{\bm r} (1+\mathcal{O}_{\bm r} \xi)^{\alpha}} .
\end{equation}
Inverting the generating function yields the PDF of $\langle \mathcal{O} \rangle_{\rm eq}$:
\begin{equation}
f_\alpha (\mathcal{O})=-\frac{1}{\pi} \lim_{\epsilon\to 0} {\rm Im} 
\frac{\sum_{\bm r} (\mathcal{O} - \mathcal{O}_{\bm r} + i \epsilon)^{\alpha -1}}{\sum_{\bm r} 
(\mathcal{O} - \mathcal{O}_{\bm r} + i \epsilon)^{\alpha }},
\label{PDF_time_average}
\end{equation}
for $\alpha <1$. In general, this PDF is not the delta function \cite{Rebenshtok2007, Rebenshtok2008}. Therefore, these observables 
depend strongly on the disorder realization, and are thus non-SA. 

For $\alpha >1$, a similar calculation for the case $\alpha <1$ gives 
\begin{equation}
f_\alpha (\mathcal{O})= \delta(\mathcal{O} - \langle \langle \mathcal{O} \rangle_{\rm eq} \rangle_{\rm dis}) \quad (L\to \infty)
\end{equation}
where we use an approximation, $\hat{\psi}_\alpha (s) \cong 1 - \langle \tau \rangle s$. 
Therefore, these observables do not depend on the disorder realization in the limit of $L\to \infty$. Hence, they have the 
 SA property for $\alpha >1$.

Using the generating function, one can obtain the moments. The first moment is given by
\begin{equation}
\langle \langle \mathcal{O} \rangle_{\rm eq} \rangle_{\rm dis} = \frac{1}{L^d} \sum_{\bm r} \mathcal{O}_{\bm r}.
\end{equation}
For $\alpha<1$, the variance has the following general relation:
\begin{equation}
\langle \langle \mathcal{O} \rangle_{\rm eq}^2 \rangle_{\rm dis} - \langle \langle \mathcal{O} \rangle_{\rm eq} \rangle_{\rm dis}^2
= (1-\alpha)  (\langle \mathcal{O}_{\bm r}^2 \rangle -\langle \mathcal{O}_{\bm r} \rangle^2),
\end{equation}
where 
\begin{equation}
\langle \mathcal{O}_{\bm r} \rangle \equiv \frac{1}{L^d} \sum_{\bm r} \mathcal{O}_{\bm r} ~{\rm and}~
\langle \mathcal{O}_{\bm r}^2 \rangle \equiv \frac{1}{L^d} \sum_{\bm r} \mathcal{O}_{\bm r}^2.
\end{equation}

Because disorder is homogeneous with respect to the axis, the disorder average of $\sigma_i^2$
can be represented by
\begin{align}
\langle \sigma_i^2 \rangle_{\rm dis} 
=  2d (\langle \langle r_k^2 \rangle_{\rm eq} \rangle_{\rm dis} - \langle \langle r_k \rangle_{\rm eq}^2 \rangle_{\rm dis}),
\label{sigma_dis_ave}
\end{align}
where $r_k$ is a position for the $k$th axis.
Considering a position and a squared position, i.e., $r_k$ and $r_k^2$, as position-dependent observables, 
one can obtain the moments. 
It follows that
\begin{align}
\langle \sigma_i^2 \rangle_{\rm dis} 
&= 2d\left\{ \sum_{k=1}^L \frac{k^2}{L} - (1-\alpha)\sum_{k=1}^L \frac{k^2}{L} 
- \alpha\left(\sum_{k=1}^L \frac{k}{L}\right)^2\right\} \nonumber\\
&\sim \alpha  \frac{dL^2}{6},
\label{sigma_alpha1}
\end{align}
for $L\to\infty$ and $\alpha <1$, while 
$\langle \sigma_i^2 \rangle_{\rm dis} \sim dL^2/6$, 
for $\alpha > 1$. At very low temperatures, the particle is situated in the minimum of the potential energy 
landscape, and hence  the fluctuations vanish when $\alpha \to 0$.

Moreover, the SA parameter for the position is given by
\begin{align}
\lim_{L\to \infty} {\rm SA}(L; {\bm r}) 
 &= \lim_{L\to \infty} \frac{\langle \langle {\bm r}\rangle_{\rm eq}^2 \rangle_{\rm dis}  - 
 \langle \langle {\bm r}\rangle_{\rm eq}\rangle_{\rm dis}^2}{\langle \langle {\bm r}\rangle_{\rm eq} \rangle_{\rm dis}^2 } \nonumber\\
 &=
\left\{
\begin{array}{ll}
0 &(\alpha >1)\\
\\
\dfrac{1-\alpha}{3}\quad &(\alpha \leq 1).
\end{array}
\right.
\label{SA_position}
\end{align}
Thus, the non-SA behavior of the position under confinement appears for $\alpha < 1$. 
Although we could not calculate the SA parameters for $\sigma_i^2$ and $t_c$, they will be non-SA for $\alpha < 1$ 
because the average position itself is not SA. 

\subsection{Ergodicity of the time-averaged MSD and position}

The EB parameter of the QTM with the reflecting boundary condition for the time-averaged MSD
can be calculated in the same way as in the periodic boundary case, whereas the $\Delta$ dependence of the 
moments of the displacement ${\bm r}(t' + \Delta) - {\bm r}(t')$ is different from that in the periodic boundary case. 
For small $\Delta$ particles rarely hit the boundary. Therefore,  the EB parameter  
is almost the same as Eq.~(\ref{EB_qtm}). 
Moreover,   time-averaged observables integrable with respect to the Boltzmann measure 
(which is a random measure) converge to the ensemble averages (averages according to the Boltzmann measure)
because particles can explore the whole space and sample all random potentials for finite systems of the QTM. 
Thus, the EB parameter of the QTM with a finite size will approach zero as $t \to \infty$. 


Because the particles eventually explore the whole system for a single disorder realization, the
time average of $\mathcal{O}_{\bm r}$ converges to the ensemble average with respect to the equilibrium state, e.g.,
$\overline{ {\bm r}(t)} \equiv \int_{0}^{t} {\bm r}(t') dt' /t  \to \langle {\bm r} \rangle_{\rm eq}$ and 
$\overline{ {\bm r}^2(t)} \equiv \int_{0}^{t} {\bm r}(t')^2 dt' /t  \to \langle {\bm r}^2 \rangle_{\rm eq}$ as $t\to \infty$. 
In general, the time averages of $\mathcal{O}_{\bm r}$ can be represented by the equilibrium probability:
\begin{equation}
\overline{\mathcal{O}} \to 
\sum_{\bm r} \mathcal{O}_{\bm r} P^{\rm eq}_{\bm r} = \frac{ \sum_{\bm r} \mathcal{O}_{\bm r} \tau_{\bm r}}{\sum_{\bm r} \tau_{\bm r}}
\end{equation}
for $t\to\infty$. Because we have the SA parameter for position with respect to the equilibrium distribution, 
the SA parameter for the time-averaged position defined by
\begin{align}
 {\rm SA}(t,L; {\bm r}) \equiv \frac{ \langle \overline{ {\bm r}(t)} ^2 \rangle_{\rm dis} - \langle \overline{ {\bm r}(t)}  \rangle_{\rm dis}^2}
{ \langle \overline{ {\bm r}(t)}  \rangle_{\rm dis}^2}
\end{align}
becomes 
\begin{align}
\lim_{L\to \infty} \lim_{t\to \infty} {\rm SA}(t,L; {\bm r}) 
 =
\left\{
\begin{array}{ll}
0 &(\alpha >1)\\
\\
\dfrac{1-\alpha}{3}\quad &(\alpha \leq 1).
\end{array}
\right.
\label{SA_position2}
\end{align}
This is the same as Eq.~(\ref{SA_position}); however, there we considered the SA property with respect to the
ensemble averages (thermal histories), whereas here we consider it with respect to the time averages. The results are the same 
because the process is ergodic, if we fix the system size and take the long-time limit.

\section{Conclusion}
We investigated ergodicity and the non-SA properties of diffusivity and position-dependent observables
 in the $d$-dimensional QTM with finite lattices, using  both periodic and reflecting boundary conditions.  
 The system is ergodic if the system size is finite. 
The transition from SA to non-SA behavior occurs at $\alpha =1$, i.e., $T=T_g$ for time-averaged MSD and position. 
Non-self averaging is a consequence of the breakdown of the law of large numbers for the waiting times at the sites. 
As a result, the non-SA effects lead to universal fluctuations of diffusivity; that is, the PDF of the 
diffusion coefficient follows the inverse L\'evy distribution in arbitrary dimensions. 

We also quantified the degree of the non-SA property by the SA parameter and showed a large difference from 
that in the corresponding annealed model (CTRW) and the infinite system of the QTM for arbitrary dimensions 
 (see Fig.~\ref{SA_L=1000}).  In other words, 
sample-to-sample fluctuations in the finite systems are different from trajectory-to-trajectory fluctuations in the corresponding infinite systems.
This difference  implies that  the limits for $L$ and $t$ are unexchangeable. 
For finite measurement times, the SA parameter of the QTM for a finite system is similar to the EB parameter of the QTM for an infinite system
 when the following conditions are satisfied: 
 the system and the measurement time are large enough to sample several different potentials,  
 but the measurement time is not sufficiently large for trajectories to traverse all sites (smaller than the characteristic time 
 of the coverage of the system's phase space). 
In the asymptotic limit, the SA parameter approaches the theoretical value (see Fig.~\ref{SA_t-dependence}). 
In contrast to the annealed model,  the SA parameter for  the finite size QTM is much larger than the corresponding fluctuations 
in the annealed case, especially when $\alpha$ is small.

There are many biological experiments on diffusion in heterogeneous environments, which  are considered to be  
  quenched environments \cite{Graneli2006, Wang2006, Kuhn2011}. 
In experiments so far, diffusion maps have been used to characterize the inhomogeneous system. 
The diffusivity map in the QTM becomes highly heterogeneous when $\alpha$ is smaller than one. This heterogeneity results from the random energy 
landscape because the local diffusivity is correlated with the energy (deep energy implies small diffusivity), whereas the actual system is 
more complicated; e.g., the upper parts are not flat, and the depths of an energy landscape will correlate with each other. 
This suggests that it is important to measure the sample-to-sample fluctuations in experiments because the disorder 
average hides rich heterogeneous structures. 

In 2008, Lubelski et al. pointed out that nonergodicity (found in the CTRW) mimics inhomogeneity, where the time-averaged MSDs for 
different realizations exhibit large fluctuations \cite{Lubelski2008}. Here, we have obtained universal 
distributions to describe the sample-to-sample fluctuations of the inhomogeneous system.  We have shown that starting from a thermal state 
and for a finite though large  system, the fluctuations stemming from inhomogeneity greatly exceed those obtained from the simpler annealed model. 
Thus, the annealed approaches hide rich physical behaviors that  are now quantified.  


\subsection*{Acknowledgements}
TA was partially supported by the Grant-in-Aid for Scientific Research (B) of the JSPS, Grant No. 16KT0021.
This research was supported by THE ISRAEL SCIENCE FOUNDATION (EB)
grant 1898/17. KS was supported by JSPS Grants-in-Aid for Scientific Research (No. JP25103003, JP16H02211, and JP17K05587).



\appendix

\section{Mean-field approximation}
It is difficult to obtain the exact result for the second moment  of $N_t$. Here, we use a mean-field approximation.
Instead of a quenched environment, we use an annealed one. In particular, 
we assume that the PDF of the trapping times does not depend on the position but follows a unique PDF given by Eq.~(\ref{pdf_L}). 
Therefore, $N_t$ can be described by a renewal process. This approximation will be valid for $d>2$. When $L$ is finite, 
all the moments of the trapping times are also finite. 

In equilibrium processes, the PDF of the first trapping time follows 
\begin{equation}
\psi_L^0 (\tau) = \frac{1}{L^d \mu_i} \sum_{k=1}^{L^d}  \exp(-\tau/\tau_k).
\end{equation}
Because the mean trapping time of Eq.~(\ref{pdf_L}) is $\mu_i$, $\langle N_t \rangle_{\rm eq}$ 
becomes 
\begin{equation}
 \langle N_t \rangle_{\rm eq} = \frac{t}{\mu_i}. 
 \label{Nt-exact}
\end{equation}
Note that this result is exact because the jump rate is always constant on average with the aid of the system's equilibration. 

The second moment of Eq.~(\ref{pdf_L}), $\langle \tau^2 \rangle_i$, is given by
\begin{equation}
\langle \tau^2 \rangle_i = \frac{2}{L^d} \sum_{k=1}^{L^d} \tau_k^2.
\end{equation}
Therefore, the second moment of $N_t$ can be calculated by renewal theory \cite{cox}:
\begin{equation}
 \langle N_t^2 \rangle_{\rm eq} = \frac{t^2}{\mu_i^2} + \frac{\langle \tau^2 \rangle_i - \mu^2_i}{\mu_i^3}t. 
\end{equation}
The first term is exact because the first moment of $N_t$ is exactly given by Eq.~(\ref{Nt-exact}).

%



\end{document}